\documentclass{elsart}

\usepackage[dvips]{graphicx}
\usepackage{amssymb}

\begin{document}

\newcommand{\beq}{\begin{equation}}
\newcommand{\eeq}{\end{equation}}
\newcommand{\beqn}{\begin{eqnarray}}
\newcommand{\eeqn}{\end{eqnarray}}

\newcommand{\Za}{{Z \alpha}}

\begin{frontmatter}


\title{Two-loop self-energy correction to the ground-state Lamb shift in H-like ions}
 \author[label1,label2]{V. A. Yerokhin},
 \ead{yerokhin@pcqnt1.phys.spbu.ru}
 \author[label3]{P. Indelicato}, and
 \author[label1]{V. M. Shabaev}
 \address[label1]{Department of Physics, St. Petersburg State University,
Oulianovskaya 1, Petrodvorets, St. Petersburg 198504, Russia}
 \address[label2]{Center for Advanced Studies, St. Petersburg State Polytechnical
University, Polytekhnicheskaya 29, St. Petersburg 195251, Russia}
 \address[label3]{Laboratoire Kastler-Brossel, \'Ecole Normale Sup\'erieure et
Universit\'e P. et M. Curie, Case 74, 4 place Jussieu, F-75252, Cedex 05,
France}

\begin{abstract}
The two-loop self-energy correction is evaluated to all orders in $\Za$ for
the ground-state Lamb shift of H-like ions with $Z\ge 20$, where $Z$ is the
nuclear charge number and $\alpha$ is the fine structure constant. The
results obtained are compared with analytical calculations performed within
the $\Za$-expansion.

\end{abstract}

\end{frontmatter}

%

The subject of this paper is a study of the historically most problematic
two-loop QED effect for the Lamb shift of H-like ions, the two-loop
self-energy correction. Traditionally, investigations of radiative
corrections to a given order in the fine-structure constant $\alpha$ rely on
an expansion in the parameters $Z\alpha$ and $\ln[(Z\alpha)^{-2}]$ ($Z$ is
the nuclear charge number). For the two-loop self-energy correction, the
energy shift is conveniently expressed in terms of a dimensionless function
$F(\Za)$,
\beq  \label{FalphaZ}
\Delta E  = m \left(\frac{\alpha}{\pi}\right)^2
\frac{(Z\alpha)^4}{n^3}\,F(Z\alpha)\,,
\eeq
which expansion reads
\beqn  \label{aZexp}
F(Z\alpha) &=& B_{40}+ (Z\alpha)B_{50} + (Z\alpha)^2 \Bigl[
  L^3 B_{63}
  +L^2 B_{62} +  L\,B_{61} + G_{\rm h.o.}(\Za) \Bigr]
  \,,
\nonumber \\
\eeqn
where  $L =\ln[(Z\alpha)^{-2}]$, $n$ is the principal quantum number, and the
remainder $G_{\rm h.o.}(Z\alpha)$ incorporates all higher-order terms.
Calculations of the coefficients of this expansion extended over several
decades
\cite{appelquist:70,pachucki:94,eides:95:pra,karshenboim:93:jetp,pachucki:01:pra,jentschura:03:jpa},
\beqn
 B_{40}(ns) &=& 1.409\,244\ldots\,,
 \\
 B_{50}(ns) &=& -24.2668(31)\,,
 \\
 B_{63}(ns) &=& -8/27\,,
 \\
 B_{62}(1s) &=& 16/27-(16/9) \ln 2\,,
 \\
 B_{61}(1s) &=& 49.838317 \,.
\eeqn
The expansion of the higher-order remainder $G_{\rm h.o.}$ starts with a
constant,
$
G_{\rm h.o.}(\Za) = B_{60}+ \Za (\cdots)\,.
$
A partial result for the $B_{60}$ coefficient was obtained recently by
Pachucki and Jentschura \cite{pachucki:03:prl},
\beq
B_{60} = -61.6(3) \pm 15\%\,,
\eeq
where $\pm 15\%$ refer to the possible contribution of uncalculated terms.
One of the notable features of the expansion (\ref{aZexp}) is its remarkably
slow convergence, even for low values of the nuclear charge number $Z$, which
exhibits itself by large numerical values of the expansion coefficients
$B_{ij}$. This makes clear that a direct evaluation of the two-loop
self-energy correction is highly desirable, especially when addressing
middle- and high-$Z$ ions.

\begin{figure}
\centerline{
\resizebox{\columnwidth}{!}{%
  \includegraphics{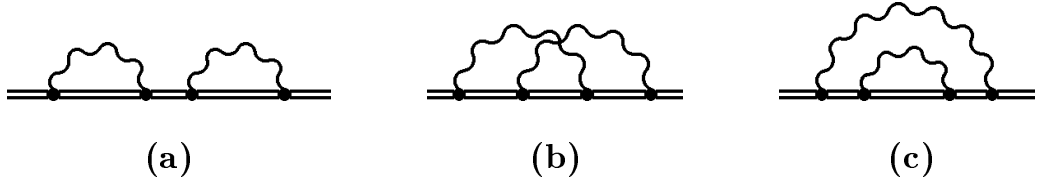}
} }
\caption{Two-loop self-energy diagrams. \label{fig:sese}}
\end{figure}

Calculations of the two-loop self-energy correction (Fig.~\ref{fig:sese}) to
all orders in the parameter $Z\alpha$ started with the irreducible
contribution of the diagram (a) (also known as the {\it loop-after-loop}
(LAL) correction), which is by far the simplest part of the total set. Such
evaluation was first accomplished by Mitrushenkov {\it et al.}
\cite{mitrushenkov:95} and later by other authors
\cite{mallampalli:98:prl,yerokhin:01:prl}. The contribution of the remaining
diagrams in Fig.~\ref{fig:sese} is by far more difficult to calculate. First
attempt to evaluate it to all orders in $Z\alpha$ was made by Mallampalli and
Sapirstein \cite{mallampalli:98:pra}. In that work, the contribution of
interest was rearranged in 3 parts, referred to by the authors as the "$M$",
"$P$", and "$F$" terms. Mallampalli and Sapirstein calculated only the $M$
and $F$ terms, while the $P$ term was left out since a new numerical
technique had to be developed for its computation. In addition, since the
numerical procedure turned out to be very time consuming, the actual
calculation of the $M$ term was carried out only for two ions, uranium and
bismuth. Subsequently, in the investigation by two of us
\cite{yerokhin:01:sese}, a computation of the remaining $P$ term for $Z=83$,
90, and 92 was performed, which formally accomplished the calculation of the
two-loop self-energy for these ions.

First complete evaluation of all contributions to the two-loop self-energy
correction to all orders in $\Za$ was carried out in our recent
investigations \cite{yerokhin:03:prl,yerokhin:03:epjd} for the ground-state
Lamb shift of H-like ions with $Z \ge 40$. In the present work, we extend our
calculation to the ions with $Z = 20$, $30$ and improve the accuracy of the
previous numerical results for $Z = 40$-60.

It is beyond the scope of the present paper to describe all modifications of
the numerical procedure that allowed us to obtain improved numerical results
in the low-$Z$ region, despite the occurrence of large numerical
cancellations that tend to grow very fast as $Z$ decreases. We only mention
that the new {\it dual-kinetic-balance} basis set \cite{shabaev:04:dkb}
constructed with $B$ splines was employed in our calculations of the $P$
term, which considerably improved the convergence with respect to the number
of basis functions in the low-$Z$ region. Another problem that was
encountered in the evaluation of the $P$ term was to keep numerical
integrations over momentum variables well under control. This problem is
associated with a significant contribution coming from the region of very
large momenta, where the numerical Green function is not smooth enough, due
to restrictions of a finite-basis-set representation. The problem was solved
by introducing a set of subtractions that have the same behaviour for large
momenta as the original integrand but are easier to evaluate numerically, and
by employing very fine grids for numerical momentum integrations. The
calculation of the $M$ term was carried out employing the contour $C_{LH}$
for the integrations over the virtual photon energies (see
Ref.~\cite{yerokhin:03:epjd}) that is much more suitable for the numerical
evaluation in the low-$Z$ region than the integration simply along the
imaginary axis. In addition, it turned out to be possible to separate from
the $M$ term a contribution, which contains the dominant part of the spurious
terms in the low-$Z$ region that are cancelled in the final sum, and to
calculate it separately. This contribution contains only one free
partial-wave expansion parameter, which makes the corresponding numerical
computation much easier as compared to the full $M$ term that has a
partial-wave expansion over two independent parameters.

Numerical results obtained for the two-loop self-energy correction for the
ground state of H-like ions are presented in Table~\ref{tab:sese} in terms of
the function $F(Z\alpha)$ defined by Eq.~(\ref{FalphaZ}). As compared to our
previous investigations \cite{yerokhin:03:prl,yerokhin:03:epjd}, we performed
calculations for $Z = 20$, 30 and improved the numerical accuracy of the
results for the $P$ and $M$ terms for $Z = 40$, 50, and 60. The values for $Z
\ge 70$ are identical to those in our previous evaluation
\cite{yerokhin:03:epjd}.

\begin{figure}
\centerline{
\resizebox{0.85\columnwidth}{!}{%
  \includegraphics{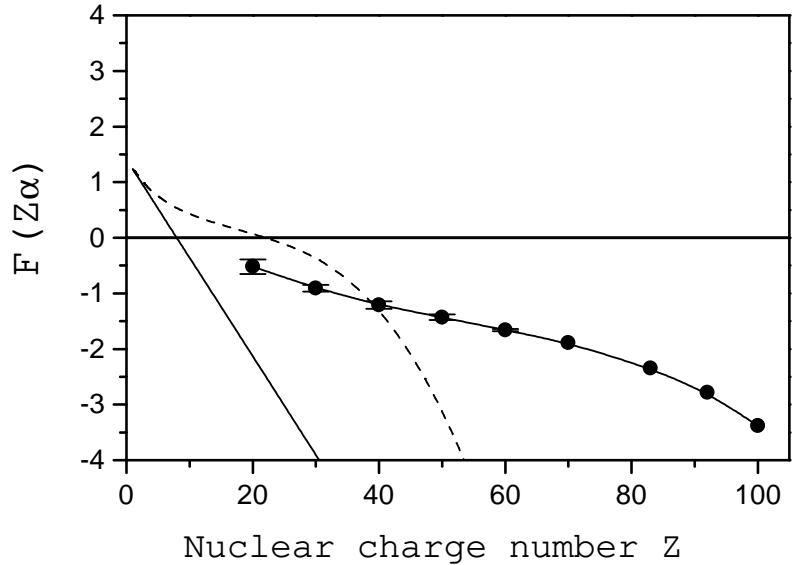}
} } \caption{The two-loop self-energy correction to all orders in $Z\alpha$
(dots) together with the contribution of the first two terms of the $Z\alpha$
expansion (\ref{aZexp}) (solid line) and all known terms of the $Z\alpha$
expansion (dashed line).  \label{fig:FZa}}
\end{figure}

\begin{figure}
\centerline{
\resizebox{0.85\columnwidth}{!}{%
  \includegraphics{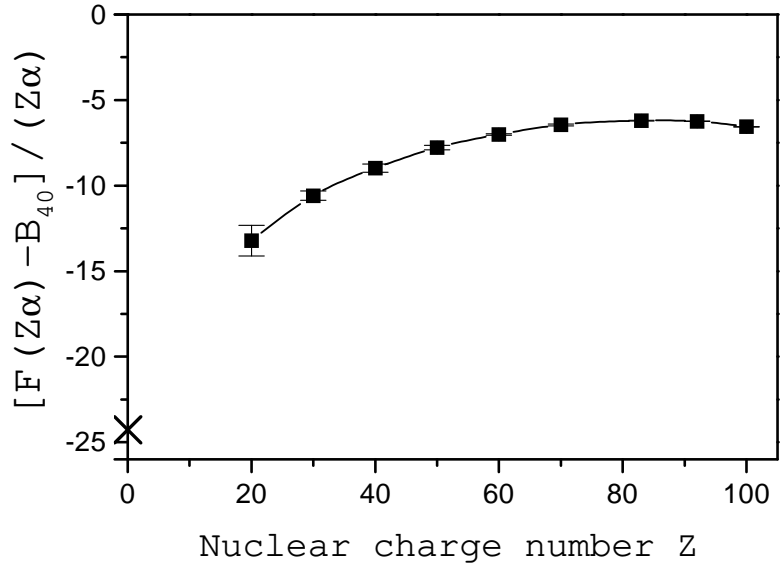}
} } \caption{Numerical results for the two-loop self-energy correction with
the leading contribution of the $\Za$ expansion subtracted out, as given by
Eq.~(\ref{BZa}). The cross denotes the analytical value for this contribution
at $Z=0$ (i.e., the coefficient $B_{50}$). \label{fig:BZa}}
\end{figure}

In Fig.~\ref{fig:FZa} we compare our non-perturbative (in $\Za$) results with
contributions of the known coefficients of the expansion (\ref{aZexp}). The
solid line corresponds to the first two terms in Eq.~(\ref{aZexp}), whereas
the dashed line represents a contribution of all known terms of the $\Za$
expansion (including the result of Ref.~\cite{pachucki:03:prl} for $B_{60}$).
In order to provide a more detailed comparison with the analytical results,
we first separate the contribution containing all orders in $\Za$ starting
from $(\Za)^2$, defined as
\beq    \label{BZa}
\widetilde{F}(\Za) = \frac{F(\Za)-B_{40}}{\Za} = B_{50}+ (\Za)(\cdots)\,.
\eeq
The function $\widetilde{F}(\Za)$ is plotted in Fig.~\ref{fig:BZa}. The cross
on the picture represents its analytical value at $Z = 0$, $B_{50}$. It is
noteworth that, despite the numerically large contributions $B_{63}$-$B_{60}$
in the next-to-leading order, the behaviour of the function
$\widetilde{F}(\Za)$ is very smooth and that it tends to the analytical value
at $Z=0$.

\begin{figure}
\centerline{
\resizebox{0.85\columnwidth}{!}{%
  \includegraphics{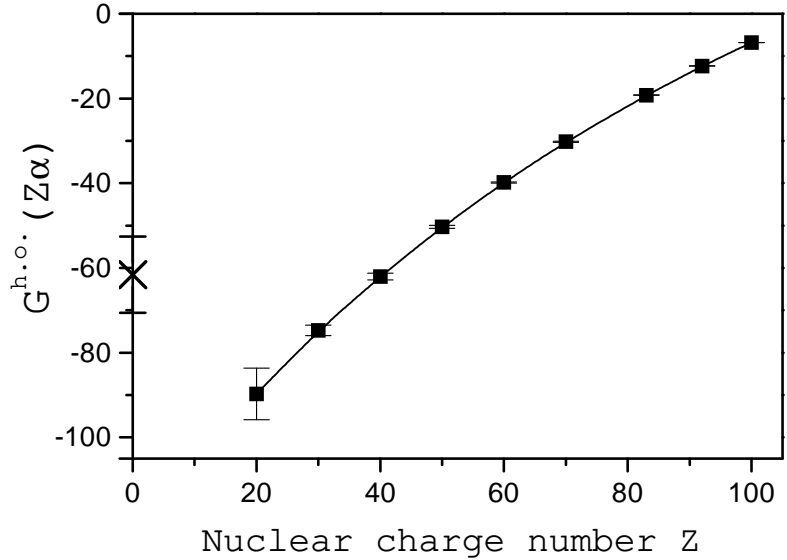}
} } \caption{Higher-order remainder $G^{\rm h.o.}(\Za)$ defined by
Eq.~(\ref{aZexp}). The cross denotes the analytical result for this
contribution at $Z=0$ (i.e., the coefficient $B_{60}$) obtained in
Ref.~\cite{pachucki:03:prl}. \label{fig:GZa}}
\end{figure}

Next, we separate the higher-order remainder $G^{\rm h.o.}(\Za)$ [defined by
Eq.~(\ref{aZexp})] from our non-perturbative results, with the corresponding
plot presented in Fig.~\ref{fig:GZa}. The cross with the error bars
corresponds to the analytical result of Ref.~\cite{pachucki:03:prl} for the
$B_{60}$ coefficient, $B_{60} = -62\pm9$. We see that the apparent limit of
the numerical values at $Z\to0$ seems to be about twice as large as the
analytical result of Ref.~\cite{pachucki:03:prl}. However, we can not at
present conclude whether this disagreement stems from the coefficient
$B_{60}$ or from one of the logarithmic coefficients in the order $\alpha^2
(\Za)^6$. A detailed comparison with individual coefficients can be in
principle performed by fitting numerical results to the analytical form of
the $\Za$ expansion, but it requires an improvement of the accuracy of the
numerical data. We mention also that a reliable extrapolation of the
numerical data to $Z=0$ is complicated by the presence of nearly-singular
terms like $\Za \ln \Za$, $\Za \ln^2 \Za$ in the $\Za$ expansion of the
remainder $G^{\rm h.o.}$, which could possibly result in a rapidly-varying
structure of  $G^{\rm h.o.}(Z)$ near the origin. We hope that an extension of
our calculations to lower values of $Z$ and an improvement of numerical
accuracy of the results will allow us to provide a more detailed comparison
with analytical calculations.

In summary, we have performed calculations of the two-loop self-energy
correction that extend our previous evaluations to the region $Z \ge 20$. The
numerical results obtained are compared with the known terms of the $\Za$
expansion. A good agreement is demonstrated with the first terms of this
expansion. However, the numerical results are shown to be likely in a
disagreement with the analytical results to order $\alpha^2 (\Za)^6$.

Valuable discussions with K. Pachucki and U. Jentschura are gratefully
acknowledged. This work was supported by INTAS YS grant No. 03-55-1442, by
RFBR grant No. 04-02-17574, and by foundation "Dynasty". The computation was
partly performed on the CINES and IDRIS national computer centers.



\begin{thebibliography}{10}
\expandafter\ifx\csname url\endcsname\relax
  \def\url#1{\texttt{#1}}\fi
\expandafter\ifx\csname urlprefix\endcsname\relax\def\urlprefix{URL }\fi

\bibitem{appelquist:70}
T. Appelquist and S.~J. Brodsky, Phys. Rev. A 2 (1970) 2293 -- 2303.

\bibitem{pachucki:94}
K.~Pachucki, Phys.  Rev. Lett. 72 (1994) 3154 -- 3157.

\bibitem{eides:95:pra}
M.~I. Eides, V.~A. Shelyuto, Phys. Rev. A 52 (1995) 954  -- 961.

\bibitem{karshenboim:93:jetp}
S.~G. Karshenboim,  Zh. Eksp. Teor. Fiz. 103 (1993) 1105 -- 1117 [JETP 76
(1993) 541 -- 546].

\bibitem{pachucki:01:pra}
K.~Pachucki, Phys. Rev. A 63 (2001) 042503--1 -- 42503--8.

\bibitem{jentschura:03:jpa}
U.~D. Jentschura,  J. Phys. A 36 (2003) L229 -- L236.

\bibitem{pachucki:03:prl}
K.~Pachucki, U.~D. Jentschura,  Phys. Rev. Lett. 91 (2003) 113005--1 --
113005--4.

\bibitem{mitrushenkov:95}
A.~Mitrushenkov, L.~Labzowsky, I.~Lindgren, H.~Persson, S.~Salomonson,  Phys.
Lett. A200 (1995) 51 -- 55.

\bibitem{mallampalli:98:prl}
S.~Mallampalli, J.~Sapirstein,  Phys. Rev. Lett. 80 (1998) 5297 -- 5300.

\bibitem{yerokhin:01:prl}
V.~A. Yerokhin,  Phys. Rev. Lett. 86 (2001)  1990 -- 1993.

\bibitem{mallampalli:98:pra}
S.~Mallampalli, J.~Sapirstein, Phys. Rev. A 57 (1998) 1548 -- 1564.

\bibitem{yerokhin:01:sese}
V.~A. Yerokhin, V.~M. Shabaev,  Phys. Rev. A 64 (2001) 062507--1 --
062507--13.

\bibitem{yerokhin:03:prl}
V.~A. Yerokhin, P.~Indelicato, V.~M. Shabaev,  Phys. Rev. Lett. 91 (2003)
073001--1 --  073001--4.

\bibitem{yerokhin:03:epjd}
V.~A. Yerokhin, P.~Indelicato, V.~M. Shabaev, Eur. Phys. J. D 25 (2003) 203
-- 238.

\bibitem{shabaev:04:dkb}
V.M. Shabaev, I.I. Tupitsyn, V.A. Yerokhin, G. Plunien, and G. Soff, Dual
kinetic balance approach to basis-set expansions for the Dirac equation,
Phys. Rev. Lett., in press.

\end{thebibliography}

%
%
\begin{table}
\caption{Individual contributions to the two-loop self-energy correction
expressed in terms of $F(Z\alpha)$.  \label{tab:sese}}
\begin{center}
\begin{tabular}{lr@{}lr@{}lr@{}lr@{}lr@{}l}
\hline
$Z$  &  \multicolumn{2}{c}{LAL}
              &  \multicolumn{2}{c}{$F$ term}
                             &  \multicolumn{2}{c}{$P$ term}
                                      &  \multicolumn{2}{c}{$M$ term}
                                               &  \multicolumn{2}{c}{Total}
\\ \hline
 20 &  $-$0&.602  &   136&.89(3)   & $-$102&.03(9)       & $-$34&.77(9)       &   $-$0&.52(13)\\
 30 &  $-$0&.757  &   44&.723(5)   &  $-$29&.42(4)       & $-$15&.46(4)       &   $-$0&.91(6) \\
 40 &  $-$0&.871  &   19&.504(2)   &  $-$11&.58(2)       &  $-$8&.26(7)       &   $-$1&.21(7)\\
    &      &      &     &          &  $-$11&.41(15)$^a$  &  $-$8&.27(18)$^a$  &   $-$1&.05(23)$^a$ \\
 50 &  $-$0&.973  &   10&.025(2)   &   $-$5&.486(20)     &  $-$5&.00(5)       &   $-$1&.43(5)\\
    &      &      &     &          &   $-$5&.41(8)$^a$   &  $-$4&.99(6)$^a$   &   $-$1&.34(10)$^a$ \\
 60 &  $-$1&.082  &    5&.723(1)   &   $-$2&.961(20)     &  $-$3&.342(9)      &   $-$1&.66(2) \\
    &      &      &     &          &   $-$2&.93(4)$^a$   &  $-$3&.342(21)$^a$ &   $-$1&.63(4)$^a$  \\
 70 &  $-$1&.216  &    3&.497      &   $-$1&.757(25)     &  $-$2&.412(11)     &   $-$1&.888(27)\\
 83 &  $-$1&.466  &    1&.938      &   $-$1&.057(13)     &  $-$1&.764(4)      &   $-$2&.349(14)\\
 92 &  $-$1&.734  &    1&.276      &   $-$0&.812(10)     &  $-$1&.513(3)      &   $-$2&.783(10)\\
100 &  $-$2&.099  &    0&.825      &   $-$0&.723(7)      &  $-$1&.384(3) &
$-$3&.381(8) \\ \hline
\end{tabular}
\end{center}
$^a\ $  Ref. \cite{yerokhin:03:epjd}.
\end{table}

\end{document}